%% file: main.tex
\documentclass[sigconf]{acmart}
\usepackage[T1]{fontenc}
\usepackage{booktabs} 
\usepackage{multirow}
\usepackage[utf8]{inputenc}
\usepackage{url}
\usepackage{graphicx}
\usepackage{color}
\usepackage{tablefootnote}
\usepackage{footnote}
\usepackage{balance}
\usepackage{tabularx}
\usepackage{textcomp}
\usepackage{caption}
\usepackage{subcaption}
\usepackage{footnote}
\usepackage{enumitem}

\usepackage{multirow}
\usepackage{amsmath}
\definecolor{Gray}{gray}{0.9}
\usepackage{colortbl}
\usepackage{balance}
\usepackage{graphicx}  
\usepackage{booktabs} 
\usepackage{multirow}
\usepackage{tabularx}
\usepackage{url}
\usepackage{hyphenat}
\usepackage{balance}
\usepackage{caption}
\usepackage{subcaption}
 \usepackage{epsfig}

\AtBeginDocument{%
  \providecommand\BibTeX{{%
    \normalfont B\kern-0.5em{\scshape i\kern-0.25em b}\kern-0.8em\TeX}}}

\copyrightyear{} 
\acmYear{} 
\setcopyright{none}
\acmConference{*This is a preprint version of a submitted paper on WebMedia. Please}{}{consider to cite the conference version instead of this one.}
\acmPrice{}  
\acmDOI{}
\acmISBN{bla} 

\begin{document}

\title{Finding Fake News Websites in the Wild*}

\author{Leandro Araujo} 
\affiliation{%
  \institution{Universidade Federal de Minas Gerais}
  \city{Belo Horizonte}
\country{Brazil}
}
\email{leandroaraujo@dcc.ufmg.br}

\author{João M. M. Couto}
\affiliation{%
  \institution{Universidade Federal de Minas Gerais}
  \city{Belo Horizonte}
\country{Brazil}
}
\email{joaocouto@dcc.ufmg.br}

\author{Luiz Felipe Nery}
\affiliation{%
  \institution{Universidade Federal de Minas Gerais}
  \city{Belo Horizonte}
\country{Brazil}
}
\email{luiznery@dcc.ufmg.br}

\author{Isadora Rodrigues}
\affiliation{%
  \institution{Universidade Federal de Minas Gerais}
  \city{Belo Horizonte}
\country{Brazil}
}
\email{isadoracmr4@gmail.com}

\author{Jussara M. Almeida}
\orcid{0000-0001-9142-2919}
\affiliation{%
  \institution{Universidade Federal de Minas Gerais}
  \city{Belo Horizonte}
\country{Brazil}
}
\email{jussara@dcc.ufmg.br}

\author{Julio C. S. Reis} 
\orcid{0000-0003-0563-0434}
\affiliation{%
  \institution{Universidade Federal de Viçosa}
  \city{Viçosa}
\country{Brazil}
}
\email{jreis@ufv.br}

\author{Fabricio Benevenuto}
\orcid{0000-0001-6875-6259}
\affiliation{%
  \institution{Universidade Federal de Minas Gerais}
  \city{Belo Horizonte}
\country{Brazil}
}
\email{fabricio@dcc.ufmg.br}

\renewcommand{\shortauthors}{Araujo, et al.}

\begin{abstract}
\input{src/00-abstract}
\end{abstract}



\maketitle

\input{src/01-introduction}
\input{src/02-related}
\input{src/03-method}

\input{src/04-validation}
\input{src/05-results}
\input{src/06-conclusion}

\begin{acks}
This work was partially supported by grants from MPMG, CNPQ, FAPEMIG, and FAPESP.
\end{acks}

\bibliographystyle{ACM-Reference-Format}
\balance
\bibliography{references}
\balance

\end{document}

%% file: src/00-abstract.tex
The battle against the spread of misinformation on the Internet is a daunting task faced by modern society. Fake news content is primarily distributed through digital platforms, with websites dedicated to producing and disseminating such content playing a pivotal role in this complex ecosystem. Therefore, these websites are of great interest to misinformation researchers. However, obtaining a comprehensive list of websites labeled as producers and/or spreaders of misinformation can be challenging, particularly in developing countries. In this study, we propose a novel methodology for identifying websites responsible for creating and disseminating misinformation content, which are closely linked to users who share confirmed instances of fake news on social media. We validate our approach on Twitter by examining various execution modes and contexts. Our findings demonstrate the effectiveness of the proposed methodology in identifying misinformation websites, which can aid in gaining a better understanding of this phenomenon and enabling competent entities to tackle the problem in various areas of society.

%% file: src/01-introduction.tex
\section{Introduction} \label{sec:introduction}

In recent times, society has been faced with an unprecedented scale of misinformation campaigns, covering highly sensitive topics including vaccines~\cite{loomba2021measuring}, climate change~\cite{treen2020online}, scientific information~\cite{west2021misinformation}, and politics \cite{reis2023helping}. The negative effects of misinformation campaigns are numerous, as they undermine the key processes used to acquire and share information, posing a significant challenge for society as a whole. Addressing this issue has become part of our daily lives, and must be tackled.

In the fight against misinformation, the complexity of the problem has emerged as one of the greatest challenges faced by modern society. The issue manifests itself in any digital platform where users consume or exchange information, including video platforms \cite{hussein2020measuring}, social networks \cite{grinberg2019fake, bovet2019influence, guess2019less}, messaging applications \cite{Resende-WWW2019,Hoseini_IMC2020}, dedicated websites, blogs, and forums \cite{setty2020truth}. The complexity of the misinformation ecosystem is further compounded by content recommendation algorithms employed by many of these platforms, which often prioritize user engagement over the accuracy of the information presented \cite{kulshrestha2015characterizing}. These factors can give rise to echo chambers, leading to polarization \cite{garimella2017long} and even the radicalization of users \cite{ribeiro2020auditing}. Additionally, social platforms allow advertisers to target users based on detailed behavioral information, allowing misinformation campaigns to target specific and sometimes vulnerable segments of a population \cite{ribeiro2019microtargeting}.

One of the key features of this intricate ecosystem is the utilization of websites dedicated to the production and dissemination of fabricated news content. These sites meticulously mimic the appearance and function of conventional and dependable news outlets. When intertwined with misinformation campaigns, they frequently attempt to manipulate public opinion, propagating widespread suspicion and distrust of credible news sources. By positioning themselves as alternative, and claiming to be more trustworthy information sources, they contribute to the creation of an alternative reality where a specific narrative and world view go unchallenged. With such a strategy, misinformation campaigns can effectively influence increasingly radicalized segments of society, serving the interests of specific political entities.

Identifying these websites and distinguishing them from their credible counterparts poses one of the most daunting challenges to the misinformation research community. Despite its undeniable significance, obtaining lists of websites that are identified as fake news sites, particularly for tackling misinformation in specific countries like Brazil, is far from a trivial task. This challenge is partially explained by the fact that misinformation campaigns are often orchestrated and supported by organizations with well-defined objectives. Those who propose to publish such lists are vulnerable to intimidation, whether by digital militias\footnote{https://www.latimes.com/91910540-132.html} or through legal harassment, involving vexatious litigation and other forms of costly legal action\footnote{https://www.abraji.org.br/entenda-o-que-e-assedio-judicial (in Portuguese)}.

This study proposes a novel approach to detecting fake news websites by leveraging user behavior rather than relying solely on website characteristics. Specifically, we hypothesize that users who share instances of fake news are likely to have shared additional ones. Our methodology identifies such users, ranks additional websites shared by them based on a specific criterion, and expands the search using articles pertaining to the newly identified websites. To validate our approach, we applied it to Twitter and compared our findings with a curated list of low-credibility websites published by an established American fact-checking website. We further applied our methodology to the Brazilian misinformation ecosystem, where we identified numerous previously unknown fake news websites. Our results demonstrate that our approach performs best when using a sorting criterion that accounts for both website impact and productivity within the relevant misinformation ecosystem and when initiated with a fake news URL checked by a recognized fact-checking entity such as the International Fact-Checking Network (IFCN\footnote{\url{https://www.poynter.org/ifcn/}}). Moreover, our study shows that users identified through our methodology are indeed more likely to post instances of fake news, thereby reducing the need for manual evaluations per identification of a low-credibility portal. We anticipate that our results will contribute to a better understanding of this phenomenon and help competent entities to address the problem in various spheres of our society.

The remainder of this paper is structured as follows: In Section \ref{sec:related}, we provide a brief overview of previous approaches taken to address the issue of identifying low-credibility content and the websites that disseminate them. Section \ref{sec:methodology} outlines the proposed methodology for identifying websites that spread misinformation. We then discuss the application of this methodology on Twitter in the American context, including details on the execution process, in Section \ref{sec:twitter}. Section \ref{sec:validation} presents the findings of this execution, which are compared to the ground truth provided by a renowned American fact-checking website to assess the efficacy of our approach. In Section \ref{sec:brazilmeasures}, we perform a "field test" of the methodology in the context of misinformation in Brazil. Finally, Section \ref{sec:conclusion} concludes the paper by highlighting its contributions and outlining future directions for research.

%% file: src/02-related.tex
\section{Related Work}
\label{sec:related}

In recent times, there has been a significant body of literature that delves into various approaches to identifying websites that are involved in creating and disseminating fake news. In this section, we aim to summarize the research that is most pertinent to our methodology, with a particular focus on three key dimensions: (i) the dynamics underlying the spread of fake news; (ii) the monetization of fake news; and (iii) network aspects of domains associated with fake news.

\subsection{Fake News Spreading Dynamics}

A number of noteworthy studies have been conducted on the dynamics of fake news on social media platforms. For instance, Vosoughi \textit{et al.} \cite{vosoughi2018spread} carried out a seminal work by analyzing rumor cascades on Twitter from 2006 to 2017. Their findings reveal that fake news reached a wider audience and spread more rapidly than accurate information. More recently, Singh \textit{et al.}~\cite{singh2020understanding} investigated the spread of URLs on Twitter during the COVID-19 pandemic. They classified URLs into different categories, such as high-quality health sources, traditional news sources, and misinformation websites, which were identified as such by the Media Bias/Fact Check (MBFC)\cite{mbfc} and other similar sources. Their results indicated that the spread of news formed a network with a sub-network of high and low credible sources. The structure of the network showed that both high and low credible sources were connected to traditional news sources, which played a critical role in bridging the two groups and facilitating the spread of information from low to high quality.

\subsection{Fake News Monetization}
Bozarth and Budak~\cite{bozarth2020market} have shown that a significant number of fake news websites, extracted from a carefully curated list, receive substantial support from reputable ad servers. This observation raises the possibility that leading ad firms could potentially help combat fake news by ceasing to provide monetization services to such websites. Meanwhile, Vekaria \textit{et al.}\cite{vekaria2022inventory} investigated how misinformation websites deceive ad server policies by pooling their ad inventory with unrelated sites in order to circumvent brand safety policies. Using a curated list of misinformation websites, they showed that misinformation websites deceptively monetize their ad inventory by exploiting a complex ad supply chain. This finding suggests that monitoring ads on misinformation websites and exposing the brands that unwittingly fund them could be potential solutions. In this regard, a recent study has characterized the effectiveness of the Sleeping Giants Brazil initiative in demonetizing fake news websites\cite{ribeiro2022sleeping}.

\subsection{Network Aspects of Fake News Web Domains}
Drawing on a curated list of websites and their corresponding credibility assessments, the study conducted by Couto \textit{et al.} \cite{couto22network} leveraged computer network attributes to reveal that fake news websites exhibit a range of content-agnostic characteristics that distinguish them from their credible counterparts. Specifically, they tend to be registered more recently, operate for shorter periods, and have certificates that expire more quickly. These findings suggest that fake news websites in Brazil are often designed to be fleeting and ephemeral, allowing them to operate with greater impunity and evade detection by authorities and fact-checkers. The study also shows that fake news websites are more likely to be hosted on foreign territories, suggesting a deliberate attempt to avoid scrutiny and regulation by local authorities. The use of computer network attributes provides an efficient and scalable addition to the toolkit of researchers working to combat the spread of misinformation.

Despite their contributions in understanding different aspects of fake news websites, such as their dynamics of dissemination, monetization methods, and network characteristics, these studies provide only a narrow glimpse into the intricate ecosystem in which these websites thrive. Moreover, the majority of these studies have mainly targeted websites from the United States, despite the worldwide scope of the misinformation problem. Therefore, our research aims to supplement existing literature by introducing a methodology that can be easily applied by researchers and practitioners in diverse regions and contexts to construct their own curated lists of websites dedicated to producing and circulating fake news on the Internet. 

%% file: src/03-method.tex
\section{Proposed Methodology}
\label{sec:methodology}

\begin{figure*}[t]
    \centering
    \includegraphics[width=0.7\linewidth]{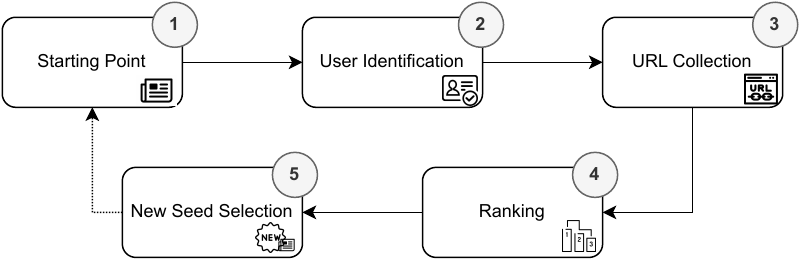}
    \caption{Overview of the proposed methodology for identifying fake news websites in the wild.}
    \label{fig:fluxo}
\end{figure*}

This section outlines the proposed methodology for detecting fake news in the wild. The underlying hypothesis is that users who have shared news articles confirmed to be instances of fake news or misinformation, based on the verdict provided by an internationally recognized fact-checking agency, are more likely to have done so on other occasions compared to users who have not. While this may not always be the case, exploring the timelines of such users, moving from one user to another based on their mutual shared content, is expected to effectively navigate the space of shared fake news articles in a given social network and identify websites closely associated with them. Our methodology consists of five main steps, as illustrated in Figure \ref{fig:fluxo}.
\begin{enumerate}
    \vspace{0.2cm}
    \item \textbf{Starting point:} The proposed methodology begins by identifying a single "seed" news article URL that is associated with misinformation content, i.e., fake news. This can be accomplished in two ways: (i) by identifying a fact-check that directly disproves information or claims made in the article, thereby establishing the article as a fake news instance, or (ii) by determining that the article was published by a low credibility website. In both cases, the credibility label must have been produced by an internationally recognized fact-checking agency such as IFCN to ensure the reliability of the seed URL.

    \vspace{0.2cm}
    \item \textbf{User identification:} The next step is to identify users who have shared the seed news article on a social media platform of choice. To validate the methodology, this study explores Twitter, which is well-suited to the target phenomenon of users sharing news articles and offers an API\footnote{\url{https://developer.twitter.com/}} that enables easy retrieval of these users' timelines.

    \vspace{0.2cm}
    \item \textbf{URL collection:} All publicly available posts made by the users identified in Step 2 are collected. From these posts, URLs are identified and extracted. The URLs are then filtered by removing those that belong to websites known not to host news articles from external sources. Examples of such websites include social networks and government websites.

    \vspace{0.2cm}
    \item\textbf{Ranking:} The websites hosting the filtered URLs are ranked according to a measure of relevance that captures their importance among the fake news-sharing users identified in Step 2. In this study, the H-Index \cite{bornmann2007we} is proposed as the metric of relevance. The websites are treated as authors, their news article URLs are treated as publications, and a user sharing one of those publications is considered a citation. The URLs contained within each of these websites are also ranked according to their importance, which is measured by a total count of shares by the users identified in Step 2.

    \vspace{0.2cm}
    \item\textbf{New seed selection:} The top-ranked news article URLs associated with the top-ranked websites are presented as candidates for addition as seeds. In this step, the URLs included in a given website's H-Index set are presented in accordance with the website standings. For instance, if the top-ranked website has an H-Index of 4, the candidates are its top-4 shared URLs. Once the entity executing the methodology selects a new seed URL, Steps 2 through 5 are repeated. This loop is referred to as a \textit{cycle}. Note that once a website's URL has been picked as a new seed on a given cycle, other URLs from the same website are no longer considered for addition in future cycles. This ensures that the methodology discovers a distinct set of news websites with each cycle.
\end{enumerate}

During the execution of the proposed methodology, at the end of each cycle, a list of new websites can be generated. This list is composed of the websites associated with the URLs selected as seeds throughout the execution cycles. This list of websites constitutes the final output of the methodology. It is important to note that the authors do not intend to release a public list of websites obtained through this methodology. The purpose of this methodology is not to accuse any website of spreading misinformation, but rather to provide a proven idea that enables researchers and competent bodies to effectively navigate fake news ecosystems by identifying websites that are closely associated with users who act as vectors for this type of content. Additionally, the proposed methodology facilitates novel research and misinformation prevention by enabling researchers to obtain their own lists of suspicious websites, which can then be evaluated for factual accuracy through fact-checks conducted by internationally recognized agencies\footnote{https://www.poynter.org/ifcn/} or agencies that assess the factuality of websites as a whole. In the following section, we present a strategy to validate the proposed methodology.

%% file: src/04-validation.tex
\section{Validation Strategy}
\label{sec:validation}

To the best of our knowledge, this methodology is unlike previously proposed approaches in the existing literature. To validate its capabilities, we investigate the effective of it at finding fake news websites, validate premise that a user that has posted a fake news instance likely posts additional ones. Also, we analyze how well the proposed methodology hold true as additional cycles are run and last, investigate if is H-Index capable of properly ranking suspicious websites for analysis, as described next. 

\subsection{Finding Ground Truth}

The validation process of our methodology poses a significant challenge, which is the establishment of a ground truth for comparison against the obtained results. In Brazil, the absence of a curated list of active fake news websites of sufficient size hinders the feasibility of convincing analysis. Conversely, in the United States, the Media Bias/Fact Check \cite{mbfc} (henceforth, MBFC) offers a potential solution. MBFC is an autonomous website that assigns high, medium, and low credibility labels, as well as political leanings, to a range of news outlets operating in the US and beyond. MBFC identifies as an independent online media outlet ``devoted to educating the public on media bias and deceptive news practices''. Although its assessments are less comprehensive than those of NewsGuard \cite{newsguard}, the assigned attributes to each news source are publicly available. In this study, we utilize the factuality labels associated with the indexed websites by MBFC. 

Thus, we compiled a dataset of websites whose credibility has been evaluated by MBFC, where each website is assigned a credibility label. In this study, we analyzed MBFC's methodology and definitions in detail and considered websites with low and questionably low credibility labels as low credibility websites. All subsequent analyses presented in the forthcoming sections were carried out based on this definition. Our dataset contains evaluations of $3,510$ distinct websites. It is worth noting, however, that while the dataset covers a broad range of websites, it is limited by the dynamic nature of the fake news ecosystem, which sees the continual emergence of new fake news websites.

\subsection{Setup}
\label{subsec:validationsetup}

As previously stated (refer to Section \ref{sec:methodology}), the proposed methodology necessitates specific input parameters and definitions for its execution. Consequently, in this section, we present the configuration of our validation strategy.

\subsubsection{Sets of initial seeds}
\label{subsubsec:initialseedsets}

To validate the effectiveness of our methodology, it is crucial to examine its performance under various initial seed conditions. The choice of initial seed plays a critical role in determining the path of website discovery throughout the execution cycles. To this end, we conduct multiple executions of the proposed methodology, each time using a different initial seed. We consider three different sets of seeds for each execution: (i) news articles derived exclusively from high-credibility websites, (ii) news articles derived from an equal proportion of fake and high-credibility websites, and (iii) news articles derived exclusively from fake news websites. Seeds from sets (i), (ii), and (iii) have 0\%, 50\%, and 100\% likelihood, respectively, of originating from fake news instances, as classified by MBFC. Through this approach, we can compare the effectiveness of our methodology in navigating the social media URL sharing ecosystem when presented with actual fake news instances versus when provided with high-credibility instances, while also validating several design decisions. The subsequent sections present the findings obtained from this setup.


\subsubsection{Automated execution and experimental setup}

In order to assess the efficacy of our methodology, it is necessary to compare it against alternative baseline approaches. For instance, we can compare the ranking of websites by their total share against the proposed H-Index method. This comparison allows us to measure the effect of the methodology's design decisions on the purpose of identifying fake news websites.

However, generating enough data points to make a thorough assessment of these effects can be costly. One of the main cost factors is Step 5 in the methodology (Figure \ref{fig:fluxo}), where a human must manually select a new URL to be used as seed for the next cycle of execution. To address this issue, we propose an ``automated'' execution of the methodology: in this approach, the algorithm is fed with a random single initial seed from one of the three sets presented in the previous section (0\%, 50\%, and 100\% fake probability). From that point forward, the most shared URL from the top ranking website, as described in Step 4, is always the one added as new seed for the following steps. Although this automation is suboptimal, as humans are better equipped to identify actual fake news instances in each cycle, it provides a lower bound for the quality of results that can be obtained in real-world usage.

 \subsection{Twitter Execution}
\label{sec:twitter}

In order to assess the potential of the methodology in practical settings, we have applied it to Twitter within the experimental framework presented previously. Through this application, we generated a dataset of results that allowed us to measure the methodology's ability to identify fake news websites, as well as to determine its properties and behavior. Twitter is a widely adopted platform for discussions on a broad range of topics, and is notorious for its use as a medium for the dissemination of misinformation campaigns~\cite{bovet2019influence, grinberg2019fake}. Although users on this platform interact with each other in various ways (e.g., follow, retweet, comment), we only considered tweets posted by users in their feeds in our work. The data was limited to tweets published in 2022. Notably, publications on Twitter may contain links to external news websites, which is precisely the domain of news content explored in this effort. The steps taken for this execution are described below. It is worth mentioning that many of these steps were only implemented to facilitate the execution of the methodology at scale, in accordance with the proposed validation strategy.

\subsubsection{Selection of initial seeds} To begin the execution of our methodology (see Figure \ref{fig:fluxo}), the first step requires the identification of a seed fake news article from which novel misinformation websites may be discovered. To gather a sizable collection of initial seeds, we utilized Twitter's search feature with the query "lang:en" which is a way of filtering tweets associated exclusively with the English language. Using the query format "max-dt:YYYY-MM-DD," we extracted a sample of tweets published between January 2022 and December 2022. This process was carried out using a newly created account to prevent the results from being biased by any user activity or recommendation algorithms.

From the resulting dataset of tweets, we identified those containing URLs. Subsequently, we filtered the URLs to extract only the ones belonging to news websites labeled in MBFC, i.e., news articles originating from sources that have a factuality label.

Henceforth, the methodology was executed strictly following the protocol outlined in Section \ref{sec:methodology}. Specifically, the following steps were carried out: (i) identification of users who have tweeted the initial seed, (ii) collection of additional URLs tweeted by these users, (iii) ranking of websites associated with these URLs according to a specified criterion, and (iv) automatic addition of the most shared link of the top-ranked website to the list of ongoing seeds. It should be emphasized that, in each cycle, users who have shared previously selected seeds were also considered for future H-index calculations. To generate a sufficient quantity of data points, the automated methodology was executed up to cycle $30$, a total of $360$ times per website ranking criterion, each one under slightly different conditions, to validate our design decisions, as discussed in the following sections.

\section{Experimental Results}

To evaluate the impact of the initial seed on the results of our methodology, we conducted an analysis of the ranking quality under different scenarios. Specifically, we investigated the effect of using initial seeds with 0\%, 50\%, and 100\% probability of being fake, as well as replacing the H-Index with two alternative ranking criteria: (i) the number of shares in the most shared URL and (ii) random ranking of websites. It is worth noting that regardless of the ranking criteria used, the most shared URL of the top-ranked website was always added to the ongoing set of seeds. By observing the pattern of ranking quality across the different initial seed scenarios, we can assess the impact of the credibility nature of the initial seed on the methodology's performance.

\begin{figure*}[t]
  \centering
  \begin{subfigure}[b]{0.33\textwidth}
    \centering
    \includegraphics[width=\textwidth] {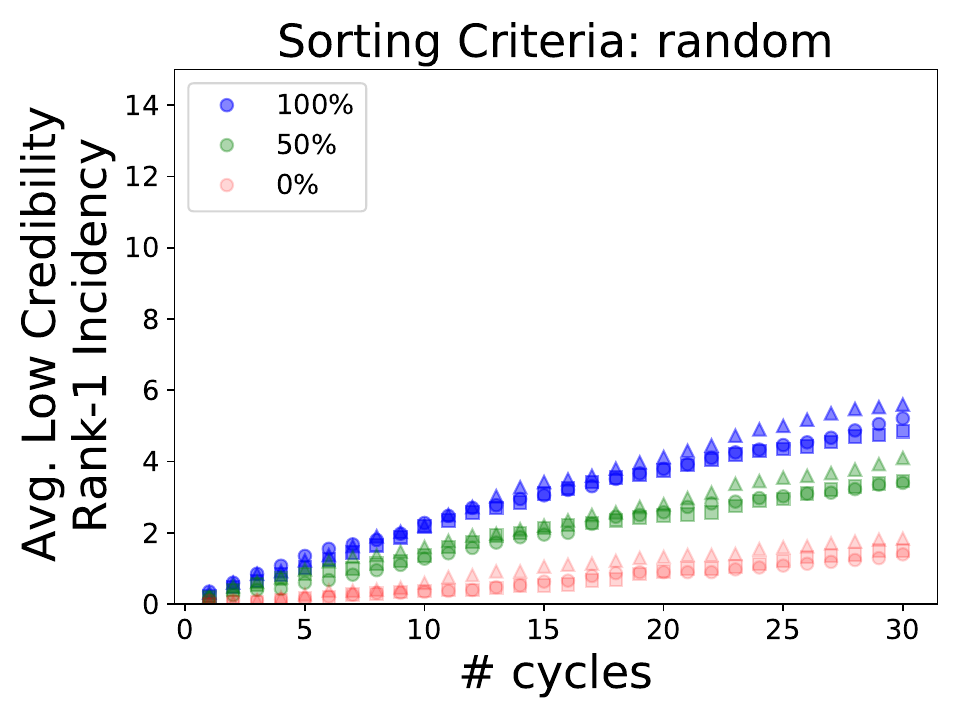}
  \end{subfigure}
  \hfill
  \begin{subfigure}[b]{0.33\textwidth}
    \centering
    \includegraphics[width=\textwidth] {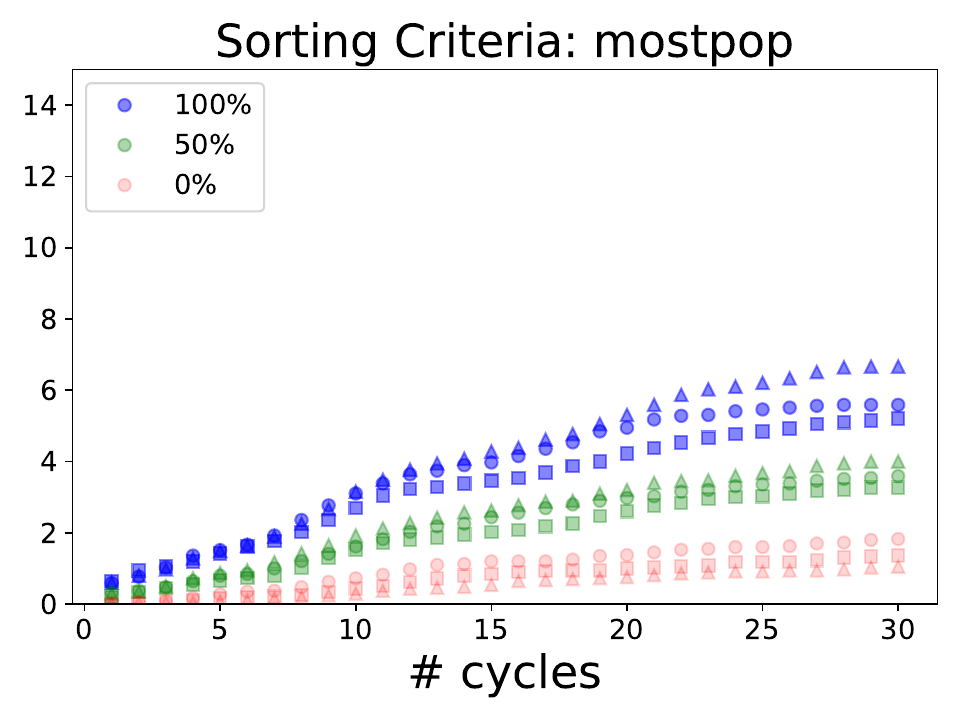}
  \end{subfigure}
  \hfill
  \begin{subfigure}[b]{0.33\textwidth}
    \centering
    \includegraphics[width=\textwidth] {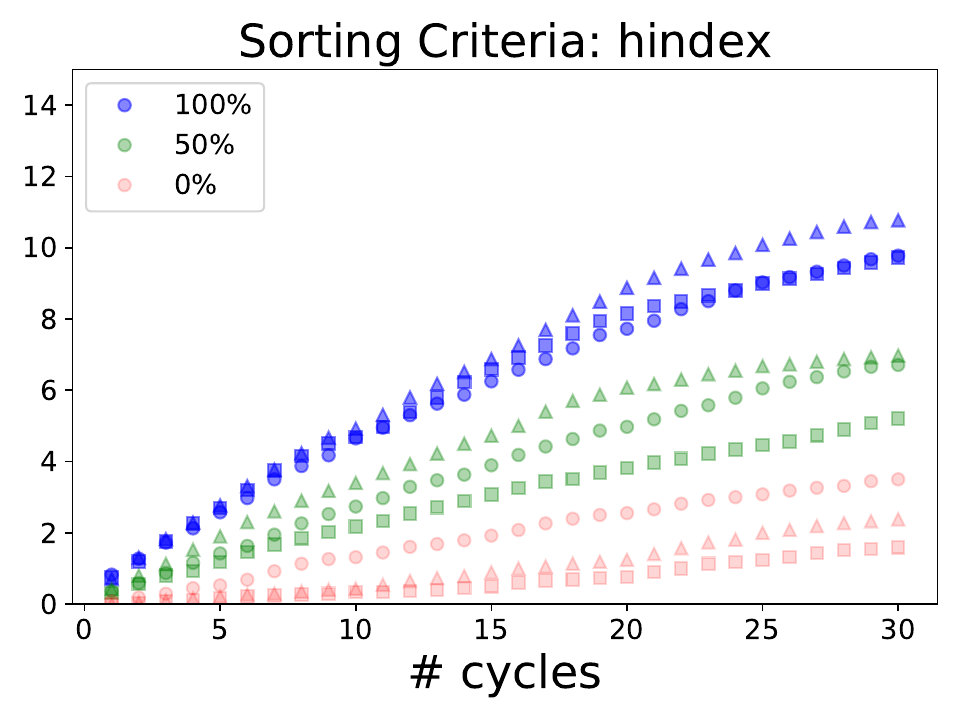}
  \end{subfigure}
  \caption{Average rank 1 incidence of low credibility websites over 40 executions with varying ranking criteria and seed dataset.}
  \label{fig:three graphs}
\end{figure*}

\subsection{Importance of the Initial Seed}
\begin{figure*}
  \centering
  \begin{subfigure}[b]{0.33\textwidth}
    \centering
    \includegraphics[width=\textwidth] {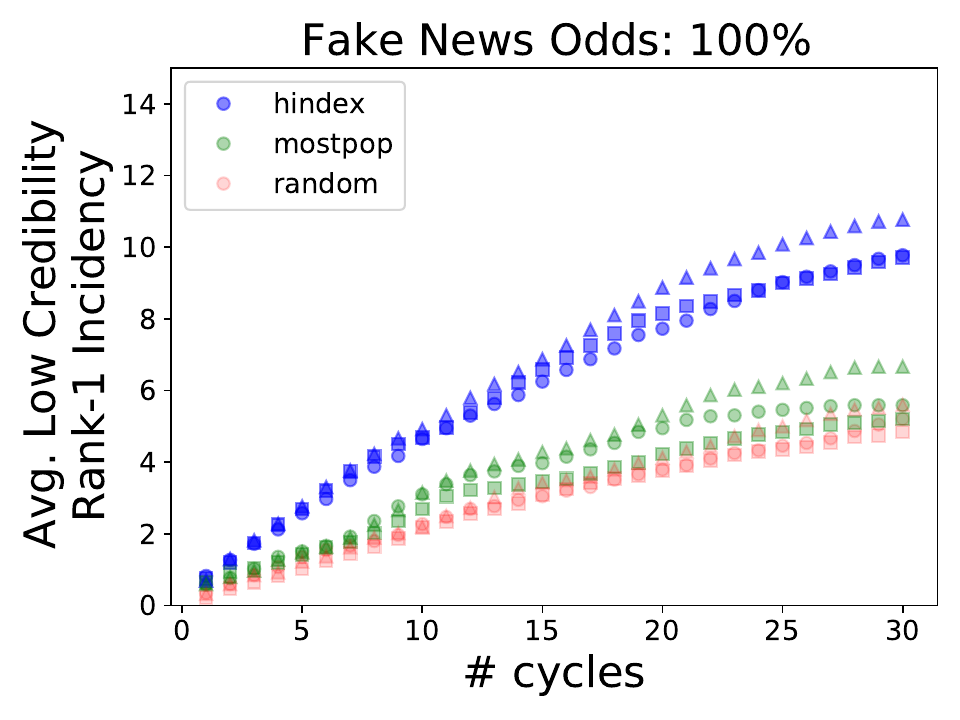}
    \caption{}
    \label{fig:incidency_prob100}
  \end{subfigure}
  \hfill
  \begin{subfigure}[b]{0.33\textwidth}
    \centering
    \includegraphics[width=\textwidth] {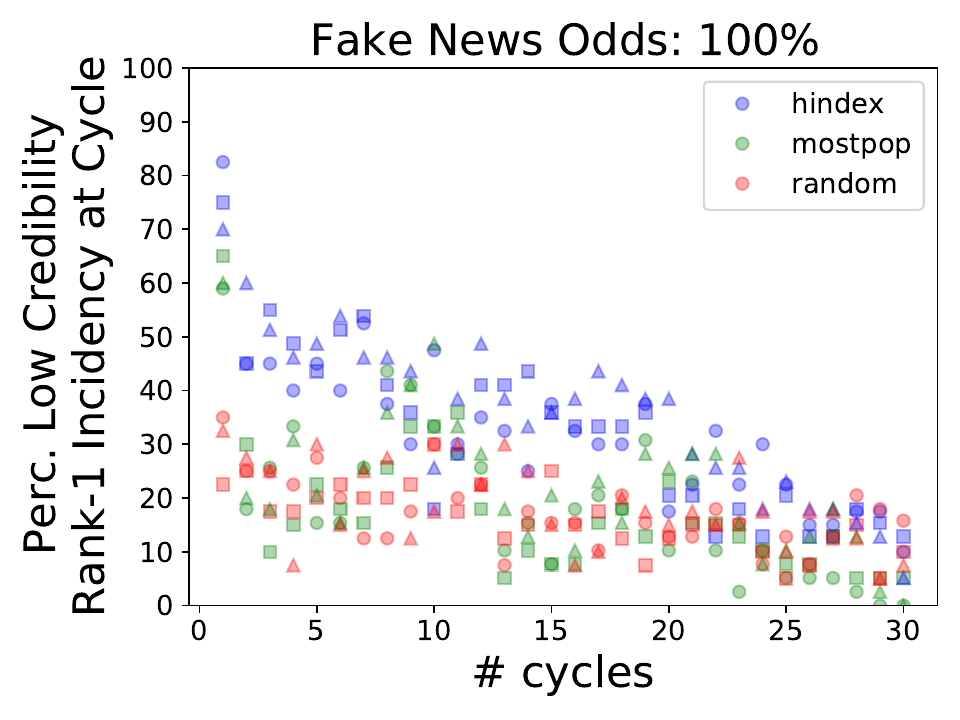}
    \caption{}
    \label{fig:density_prob100}
  \end{subfigure}
  \hfill
  \begin{subfigure}[b]{0.33\textwidth}
    \centering
    \includegraphics[width=\textwidth] {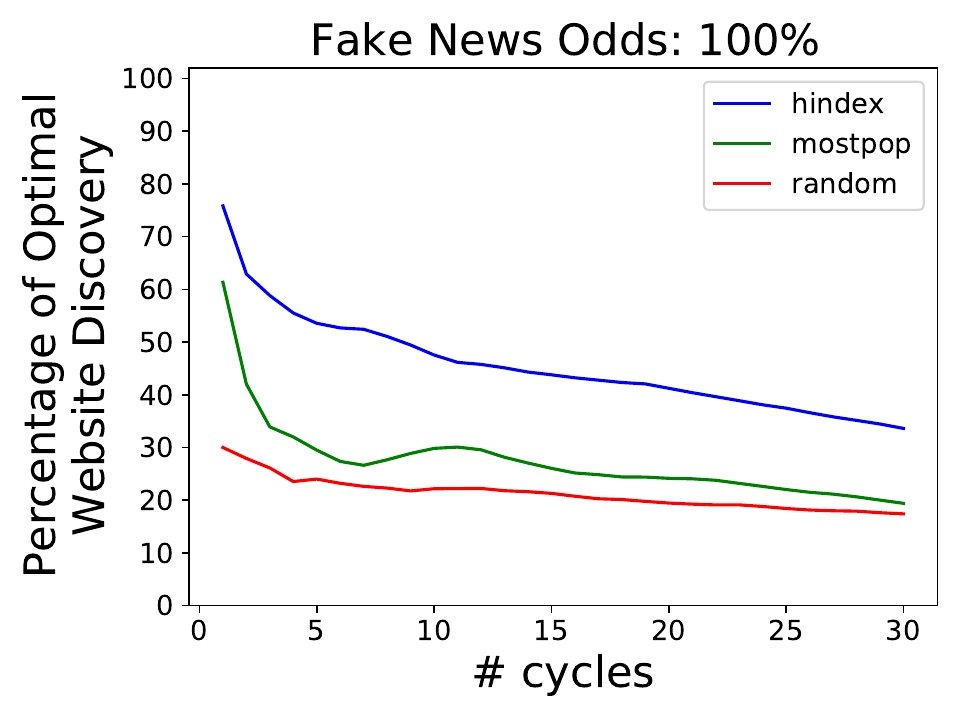}
    \caption{}
    \label{fig:recall_hindex_top1}
  \end{subfigure}
  \caption{(a) Performance for different ranking criteria when seeds are URLs of known low-credibility websites; (b) Percentage of successfully selected websites for different ranking criteria when seeds are URLs of known low-credibility websites, and; (c) Recall for different ranking criteria normalized by the best possible scenario.}
  \label{fig:odds}
\end{figure*}

The subfigures depicted in Figure \ref{fig:three graphs} present the quantity of known fake news websites per MBFC observed in the top position up to each cycle over a total of 30 cycles, each ranked by a specific ranking criterion and a varying initial set of seeds (0\%, 50\%, and 100\% fake). For each subfigure, each data point denotes the average of this quantity over 40 automated executions of each labeled scenario. For example, a data point on cycle 15 and average 10 on the 100\% curve for the ``Sorting Criteria: hindex'' graph indicates that, on average, based on the 40 executions, among the first 15 websites observed in the top-1 spot, 10 of them were fake news websites. Figure \ref{fig:incidency_prob100} presents a juxtaposition of the 100\% fake set execution curves from the aforementioned individual average fake news websites amounts graphs, so that we may better compare the differences in performance between the ranking criteria.

Finally, we conducted $3$ distinct runs of $40$ executions up to cycle $30$ for each scenario, which consists of a combination of one ranking methodology and type of initial seed, as described in Section \ref{subsec:validationsetup}. Note that each run used a unique initial seed derived from its assigned set of seeds. Notably, we found that in every cycle, even the worst of the three 40-execution runs performed under the 100\% dataset yielded a better ranking than the best performance obtained under other initial seed datasets. These results suggest that, regardless of the ranking criteria, fake news websites are consistently more likely to be ranked in the top positions throughout the cycles when the initial seed is more closely associated with misinformation. Figure \ref{fig:three graphs}, on ``Sorting Criteria: hindex'', shows that the median subset performance under 0\% probability would need 30 cycles so that 2.38 websites could be observed, against just 5 cycles under 100\% probability. This finding highlights that our methodology is capable of navigating the URL sharing landscape effectively by leveraging users with mutually shared fake news instances, resulting in the discovery of news portals of similar credibility nature to the seed.

\subsection{Website Ranking Criteria}
Having demonstrated the significance of an adequate initial seed, all subsequent analyses will be carried out assuming a 100\% fake news seed dataset, as this more closely resembles the actual implementation of our methodology, where a human is responsible for determining which URLs are added to the ongoing set of seeds, from which additional users are identified.

In Figure \ref{fig:three graphs}, we have presented a side-by-side comparison of the performance of the three different ranking criteria used in our study. The three lines in the graph represent the average number of known fake news websites observed in the top 1 position, plotted over a total of 30 cycles, with each cycle representing an execution of our methodology. As indicated in the previous paragraph, all of these executions have been performed using a 100\% fake set of seeds.

One important observation from this figure is the increasing detachment between the H-index curve and the other two curves as more cycles are completed. In other words, over time, H-index presents a significantly better ability to steer the execution towards portals that share content with a similar credibility nature as the seed. This finding is consistent with our hypothesis that H-index is a more effective metric for ranking websites in this context, as it takes into account both the popularity of the website and the diversity of the sources that link to it. The other two criteria, by contrast, are less effective at distinguishing between credible and non-credible sources.

\subsection{Fake News Website Discovery}

It is worth noting that, at the end of a cycle $x$, the methodology can, at most, have identified x fake news websites that were introduced to the set of ongoing seeds. In this regard, Figure \ref{fig:recall_hindex_top1} displays the average percentage of optimal execution achieved over the course of the cycles. Specifically, if a given cycle $x$ is linked with an average of 0.7, it implies that on cycle $x$, a total of 70\% of the $x$ websites ranked at the top position throughout the execution were fake news websites. Consequently, 0.7*$x$ fake news websites were discovered until this cycle.

The H-index curve hovering a median of approximately 50\% performance level across the 30 cycles implies that in half of the cycles, a human evaluator would have been able to identify a new low-credibility website by manually inspecting only the top-ranked website. It should be noted that this estimate represents a lower limit on the potential performance of our methodology because (1) the top-ranked website might be of low-credibility and not indexed by MBFC, and (2) human evaluators are expected to outperform our automated executions, especially if they choose seeds that are more strongly associated with misinformation than just the most shared URL from the top-ranked website. This difference arises from the fact that our automated executions are designed to replicate the real-world execution of our methodology, as described in Section \ref{subsec:validationsetup}. Thus, it is reasonable to expect better results from human evaluators in practice.

The outcome demonstrates that our proposed methodology, which incorporates H-index, commences with a low credibility seed, and restricts each website to have a single seed included throughout the cycles, enables the detection of new fake news websites without the necessity of manually verifying numerous websites before finding a single discovery. Instead, it smartly navigates the URL sharing ecosystem by leveraging the overlap of users sharing URLs associated with various fake news websites and ranks them by a metric that considers the productivity and impact of these websites. This approach results in a curated list of websites, among which a human-in-the-loop can identify new fake news websites with significantly fewer manual inspections.

\subsection{Dimishing Returns}

Figure \ref{fig:density_prob100} shows the percentage of websites ranked at the top spot that are fake news, for each individual cycle, rather than the cumulative results shown in previous graphs. Our methodology's ability to navigate the URL sharing ecosystem is reflected in the initial cycle, where the performance of the mostpop and h-index ranking criteria are similar, with both hovering around 70\% in terms of the percentage of websites on rank 1 that are fake news across the independent executions. However, as the cycles progress the performance of the different ranking criteria begins to diverge, with the h-index curve consistently outperforming the other criteria by a significant margin. By cycle 30, the performance of all ranking criteria converge to low 10's percentage, indicating the diminishing returns of the methodology as the more readily reachable websites are identified. As such, on automated executions context, it proves efficient to stop the methodology after a certain number of cycles and restart with a different set of initial seeds, highlighting the importance of a human-in-the-loop to provide feedback, keep the seeds closely related to fake news instances through the cycles and adjust the methodology to target the most promising areas for the discovery of novel fake news websites.

\subsection{Discovery of Impactful Fake News Websites}

In order to further evaluate the potential of our methodology, we aimed to assess its ability to discover the most relevant fake news websites in a given time frame. To do so, we obtained the popularity ranking for each fake news website listed in MBFC by fetching their corresponding Open Pagerank value from DomCop.org's $10$ million website dataset. The Open Pagerank score represents the importance and popularity of a website based on various factors such as the number and quality of backlinks, social media mentions, and overall web presence.

\begin{figure}[t!]
    \centering
    \includegraphics[width=0.35\textwidth] {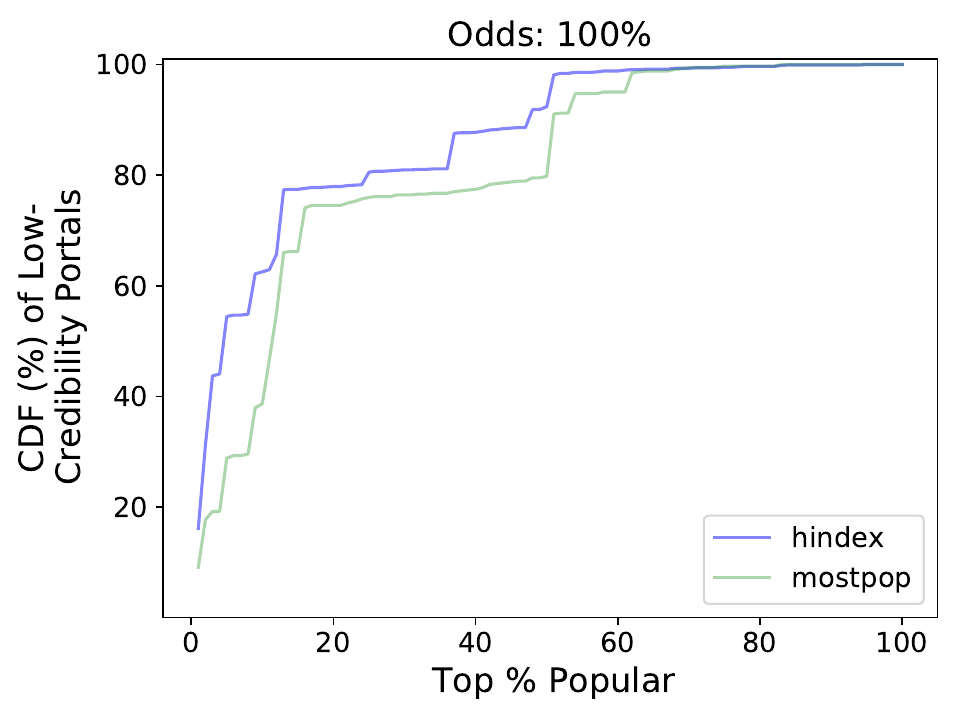}
    \caption{Cumulative distribution function (CDF) of fake news websites considering their popularity based on PageRank.}
    \label{fig:DomCop}
  \end{figure}

Using the obtained Open Pagerank scores, we constructed a cumulative distribution of the popularity rankings for the fake news websites discovered by our methodology. The resulting plot is shown in Figure \ref{fig:DomCop}. As can be seen, 60\% of the fake news websites discovered by our methodology fall within the 15\% most popular fake news websites indexed by MBFC. This indicates that our methodology by identifying and prioritizing the most influential and widely disseminated fake news websites enables more targeted and efficient interventions in the fight against spread of misinformation.

%% file: src/05-results.tex
\section{Gathering Fake News Websites in Brazil}
\label{sec:brazilmeasures}

Finally, in order to assess the generalizability of our proposed methodology, we applied it to the Brazilian context, ultimately identifying $75$ fake news websites. For this purpose, we adopted the same criteria used in our previous experiments to classify a website as a ``fake news website'', namely, if the news published on the evaluated URL was fact-checked by an internationally recognized agency and deemed to be fake. Additionally, we collected a list of $99$ news websites belonging to ANJ, the Brazilian national newspaper association responsible for defining rules and standards on news quality and factualness, in order to compare the identified fake news websites with high-credible ones. The details of our application of the methodology in the Brazilian scenario, such as the news article used as a seed and the number of cycles, are presented in another publication by our research group \cite{araujo2022identificando}.

\subsection{Relevance of Identified Websites on Social Platforms}

It is noteworthy to mention that fake news websites rely heavily on digital platforms such as Twitter and Facebook to gain traction for their publications. Consequently, it is common for them to establish an official presence on these social networks. The practicality of sharing a URL, which acts as a gateway to content hosted on an external vehicle, rather than relying on that content being available on a third-party platform, greatly enhances the websites' ability to reach a larger portion of their target audiences and establishes them as misinformation vectors worth highlighting.

In light of this, we conducted an investigation to assess the relevance of misinformation websites identified through our methodology in the context of Brazil. Specifically, we set out to find the corresponding Facebook pages of each website and measure their spread. To accomplish this, we queried the name of each website on Facebook's search feature. As a result, we were able to identify $63$ websites with a clearly corresponding Facebook page (e.g., sporting the exact same name and logo). For the remaining $13$ websites, we were unable to establish a clear mapping between them and corresponding Facebook pages.

Finally, we employed CrowdTangle\footnote{https://www.crowdtangle.com/}, a social media analytics tool, to obtain information about the popularity of each Facebook page over a period of $12$ months (from December 2021 to November 2022). We were able to retrieve CrowdTangle data for $61$ Facebook pages out of the $63$ previously mentioned, taking into consideration $5$ websites that are linked to two active pages each. Additionally, we retrieved data from $82$ pages associated with ANJ in order to compare the results. An overview of these findings is presented in Table~\ref{ct-table}.

Overall, publications by the 61 Facebook pages received $23,580,129$ shares and $160,387,038$ reactions, indicating that the identified fake news websites were able to reach approximately $30$ million users on Facebook alone. It is important to note that this number represents a lower bound for the websites identified through our proposed methodology, as we were unable to fetch the Facebook page and the CrowdTangle information for all websites. Furthermore, CrowdTangle only makes available public information about the pages, so content shared in private groups is not accounted for in our measurement. However, this experiment suggests that the set of websites discovered through the proposed methodology is highly relevant to the Brazilian fake news ecosystem.

\begin{table*}[]
\centering
\caption{Comparison of Facebook data for fake news websites that were found from the proposed methodology and credible news outlets based on ANJ.}

\begin{tabular}{|l|c|c|c|}
\hline
\textbf{Feature}        & \textbf{Fake News Websites}             & \textbf{High-Credible Sources}            & \textbf{Total}   \\ \hline
All Reactions          & 160,387,038 (43.93\%) & 204,691,475 (56.07\%) & 365,078,513   \\ \hline
Comments               & 36,188,143 (39.44\%)  & 55,558,913 (60.56\%)  & 91,747,056    \\ \hline
Likes                  & 129,658,929 (48.19\%) & 139,381,506 (51.81\%) & 269,040,435   \\ \hline
Owned Post Views       & 770,778,110 (56.62\%) & 590,538,483 (43.38\%) & 1,361,316,593 \\ \hline
Owned Total Views      & 818,999,144 (56.24\%) & 637,270,102 (43.76\%) & 1,456,269,246 \\ \hline
Owned Views from Shares & 48,221,034 (50.78\%)  & 46,731,619 (49.22\%)  & 94,952,653    \\ \hline
Page Follower Growth   & 886,397 (46.67\%)     & 1,013,009 (53.33\%)   & 1,899,406     \\ \hline
Page Followers         & 31,944,505 (34.09\%)  & 61,764,855 (65.91\%)  & 93,709,360    \\ \hline
Page Likes             & 27,119,133 (32.09\%)  & 57,402,458 (67.91\%)  & 84,521,591    \\ \hline
Shares                 & 23,580,129 (62.61\%)  & 14,084,141 (37.39\%)  & 37,664,270    \\ \hline
Total Interactions     & 220,155,313 (44.52\%) & 274,334,538 (55.48\%) & 494,489,851   \\ \hline
Total Posts            & 225,756 (23.33\%)     & 741,757 (76.67\%)     & 967,513       \\ \hline
Views on Shared Posts  & 27,241,954 (92.74\%)  & 2,131,818 (7.26\%)    & 29,373,772    \\ \hline
\end{tabular}
\label{ct-table}
\end{table*}

%% file: src/06-conclusion.tex
\section{Conclusion and Future Work}
\label{sec:conclusion}
In this study, a novel methodology for identifying websites dedicated to the production and dissemination of fake news on the internet is proposed. The approach presented is designed to be easily applicable by research and competent authorities across various geographic locations. It is worth mentioning that the current work refrains from providing a ready-made list of such websites in the interest of avoiding potential legal repercussions. Accusations directed towards specific entities regarding their involvement in the dissemination of fake news are avoided, and instead, the methodology is provided to enable research and government entities to assemble their own lists of websites. With this approach, the risk of judicial harassment is minimized, while the ability to identify fake news websites is retained. Finally, some potential research directions in this context are discussed below.

\subsection{Fake News in Different Contexts}

It is our hope that the present research serves as a catalyst for future studies on the issue of fake news worldwide. Our proposed methodology offers a comprehensive framework that enables a deeper understanding of the dissemination of misinformation on digital platforms that extends beyond Twitter analysis, including messaging apps like WhatsApp and Telegram. Our methodology takes a unique approach to this issue by targeting a common vector across all digital platforms: users sharing external fake news websites. We believe that the incidental lists generated from our method provide an opportunity for investigating various facets of fake news websites. 



\subsection{Government Action}

Our proposed methodology presents a valuable tool for regulatory entities seeking to investigate the source of funding and the beneficiaries of the publication of fake news through websites. As part of our research efforts, our group is currently engaged in a collaborative initiative with the Public Ministry of Minas Gerais (MPMG) in Brazil, providing them with sufficient material to warrant an investigation of these websites.

%% file: main.bbl

\begin{thebibliography}{25}


\ifx \showCODEN    \undefined \def \showCODEN     #1{\unskip}     \fi
\ifx \showDOI      \undefined \def \showDOI       #1{#1}\fi
\ifx \showISBNx    \undefined \def \showISBNx     #1{\unskip}     \fi
\ifx \showISBNxiii \undefined \def \showISBNxiii  #1{\unskip}     \fi
\ifx \showISSN     \undefined \def \showISSN      #1{\unskip}     \fi
\ifx \showLCCN     \undefined \def \showLCCN      #1{\unskip}     \fi
\ifx \shownote     \undefined \def \shownote      #1{#1}          \fi
\ifx \showarticletitle \undefined \def \showarticletitle #1{#1}   \fi
\ifx \showURL      \undefined \def \showURL       {\relax}        \fi
\providecommand\bibfield[2]{#2}
\providecommand\bibinfo[2]{#2}
\providecommand\natexlab[1]{#1}
\providecommand\showeprint[2][]{arXiv:#2}

\bibitem[Araujo et~al\mbox{.}(2022)]%
        {araujo2022identificando}
\bibfield{author}{\bibinfo{person}{Leandro Araujo}, \bibinfo{person}{Luiz~Felipe Nery}, \bibinfo{person}{Isadora~C Rodrigues}, \bibinfo{person}{Joao~MM Couto}, \bibinfo{person}{Julio~CS Reis}, \bibinfo{person}{Ana~PC Silva}, \bibinfo{person}{Jussara~M Almeida}, {and} \bibinfo{person}{Fabricio Benevenuto}.} \bibinfo{year}{2022}\natexlab{}.
\newblock \showarticletitle{Identificando websites de desinforma{\c{c}}ao no brasil}. In \bibinfo{booktitle}{\emph{Proc. of the Brazilian Symposium on Data Bases (SBBD)}}. \bibinfo{pages}{355--360}.
\newblock


\bibitem[Bornmann and Daniel(2007)]%
        {bornmann2007we}
\bibfield{author}{\bibinfo{person}{Lutz Bornmann} {and} \bibinfo{person}{Hans-Dieter Daniel}.} \bibinfo{year}{2007}\natexlab{}.
\newblock \showarticletitle{What do we know about the h index?}
\newblock \bibinfo{journal}{\emph{Journal of the American Society for Information Science and technology}} \bibinfo{volume}{58}, \bibinfo{number}{9} (\bibinfo{year}{2007}), \bibinfo{pages}{1381--1385}.
\newblock


\bibitem[Bovet and Makse(2019)]%
        {bovet2019influence}
\bibfield{author}{\bibinfo{person}{Alexandre Bovet} {and} \bibinfo{person}{Hern{\'a}n~A Makse}.} \bibinfo{year}{2019}\natexlab{}.
\newblock \showarticletitle{Influence of fake news in Twitter during the 2016 US presidential election}.
\newblock \bibinfo{journal}{\emph{Nature communications}} \bibinfo{volume}{10}, \bibinfo{number}{1} (\bibinfo{year}{2019}), \bibinfo{pages}{1--14}.
\newblock


\bibitem[Bozarth and Budak(2020)]%
        {bozarth2020market}
\bibfield{author}{\bibinfo{person}{Lia Bozarth} {and} \bibinfo{person}{Ceren Budak}.} \bibinfo{year}{2020}\natexlab{}.
\newblock \showarticletitle{Market forces: Quantifying the role of top credible ad servers in the fake news ecosystem}. In \bibinfo{booktitle}{\emph{The International AAAI Conference on Web and Social Media}}.
\newblock


\bibitem[Couto et~al\mbox{.}(2022)]%
        {couto22network}
\bibfield{author}{\bibinfo{person}{Joao~MM Couto}, \bibinfo{person}{Julio~CS Reis}, \bibinfo{person}{Italo Cunha}, \bibinfo{person}{Leandro Araujo}, {and} \bibinfo{person}{Fabricio Benevenuto}.} \bibinfo{year}{2022}\natexlab{}.
\newblock \showarticletitle{Characterizing Low Credibility Websites in Brazil through Computer Networking Attributes}. In \bibinfo{booktitle}{\emph{2022 IEEE/ACM International Conference on Advances in Social Networks Analysis and Mining (ASONAM)}} (Istanbul, Turkey). IEEE/ACM.
\newblock


\bibitem[Garimella and Weber(2017)]%
        {garimella2017long}
\bibfield{author}{\bibinfo{person}{Venkata Rama~Kiran Garimella} {and} \bibinfo{person}{Ingmar Weber}.} \bibinfo{year}{2017}\natexlab{}.
\newblock \showarticletitle{A long-term analysis of polarization on Twitter}. In \bibinfo{booktitle}{\emph{Eleventh international AAAI conference on web and social media}}.
\newblock


\bibitem[Gomes~Ribeiro et~al\mbox{.}(2022)]%
        {ribeiro2022sleeping}
\bibfield{author}{\bibinfo{person}{B\'{a}rbara Gomes~Ribeiro}, \bibinfo{person}{Manoel Horta~Ribeiro}, \bibinfo{person}{Virgilio Almeida}, {and} \bibinfo{person}{Wagner Meira~Jr.}} \bibinfo{year}{2022}\natexlab{}.
\newblock \showarticletitle{Analyzing the “Sleeping Giants” Activism Model in Brazil}. In \bibinfo{booktitle}{\emph{14th ACM Web Science Conference 2022}} (Barcelona, Spain) \emph{(\bibinfo{series}{WebSci '22})}. \bibinfo{publisher}{ACM}, \bibinfo{address}{NY, USA}, \bibinfo{pages}{87–97}.
\newblock
\showISBNx{9781450391917}
\urldef\tempurl%
\url{https://doi.org/10.1145/3501247.3531563}
\showDOI{\tempurl}


\bibitem[Grinberg et~al\mbox{.}(2019)]%
        {grinberg2019fake}
\bibfield{author}{\bibinfo{person}{Nir Grinberg}, \bibinfo{person}{Kenneth Joseph}, \bibinfo{person}{Lisa Friedland}, \bibinfo{person}{Briony Swire-Thompson}, {and} \bibinfo{person}{David Lazer}.} \bibinfo{year}{2019}\natexlab{}.
\newblock \showarticletitle{Fake news on Twitter during the 2016 US presidential election}.
\newblock \bibinfo{journal}{\emph{Science}} \bibinfo{volume}{363}, \bibinfo{number}{6425} (\bibinfo{year}{2019}), \bibinfo{pages}{374--378}.
\newblock


\bibitem[Guess et~al\mbox{.}(2019)]%
        {guess2019less}
\bibfield{author}{\bibinfo{person}{Andrew Guess}, \bibinfo{person}{Jonathan Nagler}, {and} \bibinfo{person}{Joshua Tucker}.} \bibinfo{year}{2019}\natexlab{}.
\newblock \showarticletitle{Less than you think: Prevalence and predictors of fake news dissemination on Facebook}.
\newblock \bibinfo{journal}{\emph{Science advances}} \bibinfo{volume}{5}, \bibinfo{number}{1} (\bibinfo{year}{2019}), \bibinfo{pages}{eaau4586}.
\newblock


\bibitem[Hoseini et~al\mbox{.}(2020)]%
        {Hoseini_IMC2020}
\bibfield{author}{\bibinfo{person}{Mohamad Hoseini}, \bibinfo{person}{Philipe Melo}, \bibinfo{person}{Manoel J{\'u}nior}, \bibinfo{person}{Fabr{\'i}cio Benevenuto}, \bibinfo{person}{Balakrishnan Chandrasekaran}, \bibinfo{person}{Anja Feldmann}, {and} \bibinfo{person}{Savvas Zannettou}.} \bibinfo{year}{2020}\natexlab{}.
\newblock \showarticletitle{Demystifying the Messaging Platforms' Ecosystem Through the Lens of {Twitter}}. In \bibinfo{booktitle}{\emph{IMC'20, 20th ACM Internet Measurement Conference}} (2020). \bibinfo{publisher}{ACM}, \bibinfo{address}{Virtual Event, USA}, \bibinfo{pages}{345--359}.
\newblock
\showISBNx{9-781-4503-8138-3}
\urldef\tempurl%
\url{https://doi.org/10.1145/3419394.3423651}
\showDOI{\tempurl}


\bibitem[Hussein et~al\mbox{.}(2020)]%
        {hussein2020measuring}
\bibfield{author}{\bibinfo{person}{Eslam Hussein}, \bibinfo{person}{Prerna Juneja}, {and} \bibinfo{person}{Tanushree Mitra}.} \bibinfo{year}{2020}\natexlab{}.
\newblock \showarticletitle{Measuring misinformation in video search platforms: An audit study on YouTube}.
\newblock \bibinfo{journal}{\emph{Proceedings of the ACM on Human-Computer Interaction}} \bibinfo{volume}{4}, \bibinfo{number}{CSCW1} (\bibinfo{year}{2020}), \bibinfo{pages}{1--27}.
\newblock


\bibitem[Kulshrestha et~al\mbox{.}(2015)]%
        {kulshrestha2015characterizing}
\bibfield{author}{\bibinfo{person}{Juhi Kulshrestha}, \bibinfo{person}{Muhammad Zafar}, \bibinfo{person}{Lisette Noboa}, \bibinfo{person}{Krishna Gummadi}, {and} \bibinfo{person}{Saptarshi Ghosh}.} \bibinfo{year}{2015}\natexlab{}.
\newblock \showarticletitle{Characterizing information diets of social media users}. In \bibinfo{booktitle}{\emph{Proceedings of the international AAAI conference on web and social media}}, Vol.~\bibinfo{volume}{9}. \bibinfo{pages}{218--227}.
\newblock


\bibitem[Loomba et~al\mbox{.}(2021)]%
        {loomba2021measuring}
\bibfield{author}{\bibinfo{person}{Sahil Loomba}, \bibinfo{person}{Alexandre de Figueiredo}, \bibinfo{person}{Simon~J Piatek}, \bibinfo{person}{Kristen de Graaf}, {and} \bibinfo{person}{Heidi~J Larson}.} \bibinfo{year}{2021}\natexlab{}.
\newblock \showarticletitle{Measuring the impact of COVID-19 vaccine misinformation on vaccination intent in the UK and USA}.
\newblock \bibinfo{journal}{\emph{Nature human behaviour}} \bibinfo{volume}{5}, \bibinfo{number}{3} (\bibinfo{year}{2021}), \bibinfo{pages}{337--348}.
\newblock


\bibitem[{Media Bias/Fact Check}(2015)]%
        {mbfc}
\bibfield{author}{\bibinfo{person}{{Media Bias/Fact Check}}.} \bibinfo{year}{2015}\natexlab{}.
\newblock
\newblock
\newblock
\shownote{\url{https://mediabiasfactcheck.com/}. Accessed 30 Nov 2022}.


\bibitem[{Newsguard}(2022)]%
        {newsguard}
\bibfield{author}{\bibinfo{person}{{Newsguard}}.} \bibinfo{year}{2022}\natexlab{}.
\newblock
\newblock
\newblock
\shownote{\url{https://www.newsguardtech.com/}. Accessed 30 Nov 2022}.


\bibitem[Reis et~al\mbox{.}(2023)]%
        {reis2023helping}
\bibfield{author}{\bibinfo{person}{Julio~CS Reis}, \bibinfo{person}{Philipe Melo}, \bibinfo{person}{Fabiano Bel{\'e}m}, \bibinfo{person}{Fabricio Murai}, \bibinfo{person}{Jussara~M Almeida}, {and} \bibinfo{person}{Fabricio Benevenuto}.} \bibinfo{year}{2023}\natexlab{}.
\newblock \showarticletitle{Helping Fact-Checkers Identify Fake News Stories Shared through Images on WhatsApp}. In \bibinfo{booktitle}{\emph{Proceedings of the 29th Brazilian Symposium on Multimedia and the Web}}. \bibinfo{pages}{159--167}.
\newblock


\bibitem[Resende et~al\mbox{.}(2019)]%
        {Resende-WWW2019}
\bibfield{author}{\bibinfo{person}{Gustavo Resende}, \bibinfo{person}{Philipe Melo}, \bibinfo{person}{Hugo Sousa}, \bibinfo{person}{Johnnatan Messias}, \bibinfo{person}{Marisa Vasconcelos}, \bibinfo{person}{Jussara Almeida}, {and} \bibinfo{person}{Fabr\'{\i}cio Benevenuto}.} \bibinfo{year}{2019}\natexlab{}.
\newblock \showarticletitle{{(Mis)Information Dissemination in WhatsApp: Gathering, Analyzing and Countermeasures}}. In \bibinfo{booktitle}{\emph{The World Wide Web Conference}} (San Francisco, CA, USA) \emph{(\bibinfo{series}{WWW '19})}. \bibinfo{publisher}{Association for Computing Machinery}, \bibinfo{address}{New York, NY, USA}, \bibinfo{pages}{818–828}.
\newblock
\showISBNx{9781450366748}
\urldef\tempurl%
\url{https://doi.org/10.1145/3308558.3313688}
\showDOI{\tempurl}


\bibitem[Ribeiro et~al\mbox{.}(2019)]%
        {ribeiro2019microtargeting}
\bibfield{author}{\bibinfo{person}{Filipe~N Ribeiro}, \bibinfo{person}{Koustuv Saha}, \bibinfo{person}{Mahmoudreza Babaei}, \bibinfo{person}{Lucas Henrique}, \bibinfo{person}{Johnnatan Messias}, \bibinfo{person}{Fabricio Benevenuto}, \bibinfo{person}{Oana Goga}, \bibinfo{person}{Krishna~P Gummadi}, {and} \bibinfo{person}{Elissa~M Redmiles}.} \bibinfo{year}{2019}\natexlab{}.
\newblock \showarticletitle{On microtargeting socially divisive ads: A case study of russia-linked ad campaigns on facebook}. In \bibinfo{booktitle}{\emph{Proceedings of the conference on fairness, accountability, and transparency}}. \bibinfo{pages}{140--149}.
\newblock


\bibitem[Ribeiro et~al\mbox{.}(2020)]%
        {ribeiro2020auditing}
\bibfield{author}{\bibinfo{person}{Manoel~Horta Ribeiro}, \bibinfo{person}{Raphael Ottoni}, \bibinfo{person}{Robert West}, \bibinfo{person}{Virg{\'\i}lio~AF Almeida}, {and} \bibinfo{person}{Wagner Meira~Jr}.} \bibinfo{year}{2020}\natexlab{}.
\newblock \showarticletitle{Auditing radicalization pathways on YouTube}. In \bibinfo{booktitle}{\emph{Proceedings of the 2020 conference on fairness, accountability, and transparency}}. \bibinfo{pages}{131--141}.
\newblock


\bibitem[Setty and Rekve(2020)]%
        {setty2020truth}
\bibfield{author}{\bibinfo{person}{Vinay Setty} {and} \bibinfo{person}{Erlend Rekve}.} \bibinfo{year}{2020}\natexlab{}.
\newblock \showarticletitle{Truth be Told: Fake News Detection Using User Reactions on Reddit}. In \bibinfo{booktitle}{\emph{Proceedings of the 29th ACM International Conference on Information \& Knowledge Management}}. \bibinfo{pages}{3325--3328}.
\newblock


\bibitem[Singh et~al\mbox{.}(2020)]%
        {singh2020understanding}
\bibfield{author}{\bibinfo{person}{Lisa Singh}, \bibinfo{person}{Leticia Bode}, \bibinfo{person}{Ceren Budak}, \bibinfo{person}{Kornraphop Kawintiranon}, \bibinfo{person}{Colton Padden}, {and} \bibinfo{person}{Emily Vraga}.} \bibinfo{year}{2020}\natexlab{}.
\newblock \showarticletitle{Understanding high-and low-quality URL Sharing on COVID-19 Twitter streams}.
\newblock \bibinfo{journal}{\emph{Journal of Computational Social Science}} \bibinfo{volume}{3}, \bibinfo{number}{2} (\bibinfo{year}{2020}), \bibinfo{pages}{343--366}.
\newblock


\bibitem[Treen et~al\mbox{.}(2020)]%
        {treen2020online}
\bibfield{author}{\bibinfo{person}{Kathie M~d'I Treen}, \bibinfo{person}{Hywel~TP Williams}, {and} \bibinfo{person}{Saffron~J O'Neill}.} \bibinfo{year}{2020}\natexlab{}.
\newblock \showarticletitle{Online misinformation about climate change}.
\newblock \bibinfo{journal}{\emph{Wiley Interdisciplinary Reviews: Climate Change}} \bibinfo{volume}{11}, \bibinfo{number}{5} (\bibinfo{year}{2020}), \bibinfo{pages}{e665}.
\newblock


\bibitem[Vekaria et~al\mbox{.}(2022)]%
        {vekaria2022inventory}
\bibfield{author}{\bibinfo{person}{Yash Vekaria}, \bibinfo{person}{Rishab Nithyanand}, {and} \bibinfo{person}{Zubair Shafiq}.} \bibinfo{year}{2022}\natexlab{}.
\newblock \showarticletitle{The Inventory is Dark and Full of Misinformation: Understanding the Abuse of Ad Inventory Pooling in the Ad-Tech Supply Chain}.
\newblock \bibinfo{journal}{\emph{arXiv preprint arXiv:2210.06654}} (\bibinfo{year}{2022}).
\newblock


\bibitem[Vosoughi et~al\mbox{.}(2018)]%
        {vosoughi2018spread}
\bibfield{author}{\bibinfo{person}{Soroush Vosoughi}, \bibinfo{person}{Deb Roy}, {and} \bibinfo{person}{Sinan Aral}.} \bibinfo{year}{2018}\natexlab{}.
\newblock \showarticletitle{The spread of true and false news online}.
\newblock \bibinfo{journal}{\emph{science}} \bibinfo{volume}{359}, \bibinfo{number}{6380} (\bibinfo{year}{2018}), \bibinfo{pages}{1146--1151}.
\newblock


\bibitem[West and Bergstrom(2021)]%
        {west2021misinformation}
\bibfield{author}{\bibinfo{person}{Jevin~D West} {and} \bibinfo{person}{Carl~T Bergstrom}.} \bibinfo{year}{2021}\natexlab{}.
\newblock \showarticletitle{Misinformation in and about science}.
\newblock \bibinfo{journal}{\emph{Proceedings of the National Academy of Sciences}} \bibinfo{volume}{118}, \bibinfo{number}{15} (\bibinfo{year}{2021}), \bibinfo{pages}{e1912444117}.
\newblock


\end{thebibliography}
